\documentclass[12pt,letterpaper]{article}
\usepackage{amssymb,amsmath}

\begin{document}
\pagestyle{plain}
\begin{center}
\vspace{1.cm}

{\Large \bf{
Curved Space (Matrix) Membranes}}\\[1cm]
{\large Jens Hoppe}
\\[.4cm]
\vspace{.2cm}
{\small \textit{ 
Department of Mathematics, 
Royal Institute of Technology, 100 44 Stockholm}}
\end{center}

\vspace{1cm}

\begin{abstract}
\centerline{\mbox{Hamiltonian formulations of M-branes moving in curved backgrounds are given.}}
\end{abstract}

As is well known (see e.g. \cite{1}, \cite{2}), varying the volume

\begin{eqnarray}
 {\rm Vol}\: \mathcal{M} = S[x^{\mu}]=\int d^{M+1}\varphi\sqrt{G} \\
 G= \left \rvert \det \left(  \dfrac{\partial x^{\mu}}{\partial \varphi^{\alpha}}
    \dfrac{\partial x^{\nu}}{\partial \varphi^{\beta}} \eta_{\mu\nu}(x)
     \right) \right \rvert  \nonumber
 \label{eq1}    
\end{eqnarray}
of an $M+1$ dimensional (time-like) manifold $\mathcal{M}$ embedded in a 
Lorentzian manifold $\mathcal{L}$ one obtains

\begin{eqnarray}
 \dfrac{1}{\sqrt{G}} \partial_{\alpha}\left( \sqrt{G} G^{\alpha\beta}
  \partial_{\beta} x^{\mu }\right) + G^{\alpha\beta}
  \partial_{\alpha} x^{\nu }\partial_{\beta} 
   x^{\lambda}\Gamma^{\mu}_{\nu\lambda}(x)=0\\
  \mu=0,1,...N \nonumber
 \label{eq2}
\end{eqnarray}
as equations of motion. Assuming that $\mathcal{L} = \mathbb{R}\times \mathcal{N}$, and choosing 
\cite{3}

\begin{eqnarray}
\label{eq3}
  \varphi^{0}=x^{0}=:t  \ , \ G_{0b}= 0 \\ 
  (b=1,...M), \nonumber
\end{eqnarray}
referred to as O(rdinary)T(ime) Orthonormal Gauge, (\ref{eq2}) \begin{scriptsize}$\mu=0$ \end{scriptsize} 
implies that

\begin{equation}
  \rho:= \left( \dfrac{\det (\partial_{a}x^{i}\partial_{b}x^{j} \eta _{ij}(x))} 
   {1-\dot{x}^{i}\dot{x}^{j}\eta_{ij}(x)}\right)^{1/2}
\label{eq4}
\end{equation}
is time-independent, and (\ref{eq2}) \begin{scriptsize}$\mu=i..N$\end{scriptsize} then 
takes the form

\begin{equation}
  \ddot{x}^{i}+\dot{x}^{j}\dot{x}^{k}\Gamma^{i}_{jk}(x)=
  \dfrac{1}{\rho}\partial_{a}\left( \dfrac{g}{\rho} g^{ab}\partial_{b}x^{i}\right) 
  + \dfrac{g}{\rho^{2}} g^{ab}\partial_{a}x^{j}\partial_{b}x^{k}\Gamma^{i}_{jk}(x).
\label{eq5}
\end{equation}
As shown below, (\ref{eq5}) -together with (\ref{eq3}) and (\ref{eq4}), 
are Hamiltonian with respect to

\begin{eqnarray}
 H = \int d^{M}\varphi\sqrt{p_{i}\eta^{ij}(x)p_{j} + 
 \det\left(\partial_{a}x^{i}\partial_{b}x^{j}\eta_{ij}(x)\right)}
 =: \int d^{M}\varphi \mathcal{H}, 
\label{eq6}
\end{eqnarray}
 
\begin{eqnarray}
 p_{i}\partial_{a}x^{i}=0 \
 (a=1...M) 
\label{eq7}
\end{eqnarray}
and\footnote{similar results were, as far as I know, also perceived by 
V. Moncrief, who tried to generalize \cite{4}, \cite{5} to curved backgrounds
, one or two years ago.}the Hamiltonian density $\mathcal{H}=\sqrt{p^2 + g}$
is time independent (and equal to $\rho$), just as in the case of flat 
backgrounds \cite{4}, \cite{5}. The equations of motion following from (\ref{eq6}) are

\begin{eqnarray}
 \dot{x}^{i} = \dfrac{\delta H}{\delta p_{i}} = \frac{\eta^{ij}(x)}{\mathcal{H}}p_{j} \\
 \dot{p}_{i} = \dfrac{-\delta H}{\delta x^{i}} = \dfrac{-1}{2\mathcal{H}}p_{j}p_{k}
 \partial_{i}\eta^{jk} - \dfrac{1}{2\mathcal{H}}g g^{ab}\partial_{a}x^{j}\partial_{b}x^{k}
 \partial_{i}\eta_{jk} \nonumber \\ + \partial_{a}\left(gg^{ab}\eta_{ij}
 \dfrac{\partial_{b}x^{j}}{\mathcal{H}}\right). \nonumber
 \label{eq8}
\end{eqnarray} 
Using (e.g.)

\begin{eqnarray}
 \partial_{c}x^{i}\partial_{a}\left( g g^{ab}\eta_{ij}
 \dfrac{\partial_{b}x^{j}}{\mathcal{H}} \right) 
  = g g^{ab}\eta_{ij} \partial_{c}x^{i} \partial_{b}x^{j} \partial_{a}
 \left( \dfrac{1}{\mathcal{H}}\right) \nonumber\\
 +\dfrac{1}{\mathcal{H}}\partial_{a}\left(g g^{ab}\eta_{ij} \partial_{c}x^{i}
   \partial_{b}x^{j}\right) - \dfrac{1}{\mathcal{H}}g g^{ab}\eta_{ij} 
 \partial_{b}x^{j}\partial_{ac}^{2}x^{i} \\
 = g \partial_{c}\left( \dfrac{1}{\mathcal{H}}\right) 
 +\dfrac{1}{\mathcal{H}}\partial_{c}g 
 -\dfrac{1}{2\mathcal{H}}\partial_{c}g  + \dfrac{1}{2\mathcal{H}}g g^{ab}
  \partial_{a}x^{j} \partial_{b}x^{k}\partial_{c}\eta_{jk}\nonumber
\label{eq9}
\end{eqnarray}
one first shows that

\begin{eqnarray}
 \partial_{t}(p_{i}\partial_{c}x^{i})= ... = \dfrac{1}{2\mathcal{H}}\partial_{c}
 \left( p_{i}\eta^{ij}p_{j} + g \right) + \left( p^2 + g\right)\partial_{c} 
 \dfrac{1}{\mathcal{H}} = 0
\label{eq10}
\end{eqnarray}
and then, using (\ref{eq7}) (three times), 

\begin{equation}
 \dot{\mathcal{H}}=0.
\label{eq11}
\end{equation}
Note that in the case of membranes, 

\begin{equation}
 g/\rho^{2}= -\dfrac{1}{2}\{x^{i},x^{k}\}\eta_{kj}\{x^{j},x^{l}\}\eta_{li},
\label{eq12}
\end{equation}
if

\begin{equation}
 \{f,h\} := \frac{\in^{ab}}{\rho}\partial_{a}f\partial_{b}h,
\label{eq13}
\end{equation}
and that, due to

\begin{equation}
 gg^{ab}=\epsilon^{aa'}\epsilon^{bb'}g_{a'b'}
\label{eq14}
\end{equation}
when $M=2$, the r.h.s. of (\ref{eq5}) may be written as

\begin{equation}
 \{\eta_{jk}\{x^{i},x^{k}\},x^{j}\}+\{x^{j},x^{l}\}\{x^{k},x^{m}\}
  \eta_{lm}\Gamma^{i}_{jk},
\label{eq15}
\end{equation}
hence allowing a matrix model approximation of (\ref{eq5}). However, 
as in the case of flat backgrounds \cite{3}, due to the constraints 
(\ref{eq7}) and the square root in (\ref{eq6}), (especially with respect to
quantization), it can be advantageous to use light-cone coordinates
and (assuming $\mathcal{M}= \mathbb{R}^{1,1}\times \tilde{\mathcal{N}}$)
choose the L(ight)C(one) ONG (cp \cite{3} for $\tilde{\mathcal{N}}=\mathbb{R}^{d}$)

\begin{eqnarray}
 \varphi^{0} = \dfrac{x^{0}+x^{d+1}}{2}, \\
 G_{0b}=0 \ (b=1...M). \nonumber
\label{eq16}
\end{eqnarray}
One again obtains (\ref{eq5}), but now with

\begin{equation}
 ij=1...N-1 =:d
\label{eq17}
\end{equation}
 and the integrability of

\begin{equation}
 0= G_{0b}= \partial_{b}\zeta - \eta_{ij}\dot{x}^{j}\left(\partial_{b}x^{i}\right), 
\label{eq18}
\end{equation}
allowing to determine $ \zeta= x^{0}-x^{d+1}$, implies

\begin{equation}
 \partial_{a}\left( \eta_{ij}\dot{x}^{j}\right)\partial_{b}x^{i} -
 \partial_{b}\left( \eta_{ij}\dot{x}^{j}\right)\partial_{a}x^{i}=0,
\label{eq19}
\end{equation}
instead of (\ref{eq7}). With

\begin{equation}
 \rho = \left( \dfrac{\det(\partial_{a}x^{i}\partial_{b}x^{j}\eta_{ij}(x))}
 {2\dot{\zeta}-\dot{x}^{i}\dot{x}^{j}\eta_{ij}} \right)^{1/2}
\label{eq20}
\end{equation}
being again a time independent, non-dynamical density, a light-cone Hamiltonian description
is then given by

\begin{equation}
  H = \dfrac{1}{2}\int \dfrac{d^{M}\varphi}{\rho}\left( p_{i}\eta^{ij}p_{j} + g \right) ,
\label{eq21} 
\end{equation}

\begin{equation}
\partial_{a}p_{i}\partial_{b}x^{i}-\partial_{b}p_{i}\partial_{a}x^{i}=0 .
\label{eq22}
\end{equation}
In particular,  for $M=2$ (see  \cite{6},\cite{7} for some related results)

\begin{equation}
  H = \dfrac{1}{2}\int d^{2}\varphi \rho \left(\dfrac{ p_{i}}{\rho}\eta^{ij}\dfrac{p_{j}}
  {\rho} - \dfrac{1}{2}\{x^{i},x^{k}\}\eta_{kj} \{x^{j}x^{l}\}\eta_{li}  \right),
\label{eq23}
\end{equation}

\begin{equation}
 \{ p_{i},x^{i}\}=0,
\label{eq24}
\end{equation}
obviously leading to a matrix model with 

\begin{eqnarray}
 H \sim \rm{Tr}\left( P_{i}\eta^{ij}(X)P_{j} + \dfrac{1}{2} \left[X^{i},X^{k}\right]
 \eta_{kj}(X)\eta_{li}(X)\left[X^{j},X^{l}\right] \right) \nonumber\\
  \sum ^{d}_{i=1}\left[ P_{i},X^{i}\right]= 0.
\end{eqnarray}

\section*{Acknowledgements}
I would like to thank Joakim Arnlind and Martin Bordemann
for discussions, and Jan Plefka for bringing
reference [6] to my attention when discussing together after my seminar at the Albert Einstein Institute.

\end{document}